\documentclass{kapproc} 

\usepackage{procps} 

\usepackage[dvips]{graphicx}

\upperandlowercase
\kluwerbib 

\begin{document}

\articletitle{The Disk Mass Project}

\articlesubtitle{breaking the disk-halo degeneracy}

\author{Marc A.W. Verheijen$^{1*}$, Matthew A. Bershady$^2$, Rob A. Swaters$^3$,\\ David R. Andersen$^4$ and Kyle B. Westfall$^2$}
\affil{$^1$Kapteyn Astronomical Institute, $^2$University of Wisconsin -- Madison,\\ $^3$University of Maryland, $^5$NRC Herzberg Institute of Astrophysics}
\email{$^*$verheyen@astro.rug.nl}


\begin{abstract}
Little is known about the content and distribution of dark matter in
spiral galaxies. To break the degeneracy in galaxy rotation curve
decompositions, which allows a wide range of dark matter halo density
profiles, an independent measure of the mass surface density of
stellar disks is needed. Here, we present our ongoing Disk Mass
project, using two custom-built Integral Field Units, to measure the
vertical velocity dispersion of stars in $\sim$40 spiral
galaxies. This will provide a kinematic measurement of the stellar
disk mass required to break the degeneracy, enabling us to determine the dark
matter properties in spiral galaxies with unprecedented accuracy. Here
we present preliminary results for three galaxies with different
central disk surface brightness levels.
\end{abstract}


\section{Motivation}

A major roadblock in testing galaxy formation models is the disk-halo
degeneracy: density profiles of dark matter halos as inferred from
rotation curve decompositions depend critically on the adopted M/L of
the disk component. An often used refuge to circumvent this degeneracy
is the adoption of the maximum-disk hypothesis (van Albada \& Sancisi
1986).  However, this hypothesis remains unproven. Bell \& De Jong
(2001) showed that stellar population synthesis models yield plausible
{\it relative} measurements of stellar M/L in old disks, but
uncertainties in the IMF prevent an {\it absolute} measurement of
stellar M/L from photometry. Another tool to determine the M/L, and
specifically whether disks are maximal, is the Tully-Fisher relation,
e.g. by looking for offsets between barred vs. un-barred galaxies, but
this too is only a relative measurement. Evidently, none of
these methods are suited to break the degeneracy, and without an
independent measurement of the M/L of the stellar disk, it is not
possible to determine the structural properties of dark matter halos
from rotation curve decompositions.

A direct and absolute measurement of the M/L can be derived from the
vertical component $\sigma_{\rm z}$ of the stellar velocity
dispersion. For a locally isothermal disk, $\sigma_{\rm z} = \sqrt{\pi
G(M/L)\mu z_o}$, with $\mu$ the surface brightness, and $z_0$ the disk
scale height. The latter is statistically well-determined from studies
of edge-on galaxies (de Grijs \& van der Kruit 1996, Kregel et
al 2002). Thus, $\sigma_{\rm z}$ provides a direct, kinematic
estimate of the M/L of a galaxy disk and can break the disk-halo
degeneracy.

This approach has been attempted before with long-slit spectroscopy on
significantly inclined galaxies. For example, Bottema (1997) concluded
for a sample of 12 galaxies that, on average, the stellar disk
contributes at maximum some 63\% to the amplitude of the rotation
curve. These observations, however, barely reached 1.5 disk
scale-lengths, required broad radial binning, and because of the high
inclinations, the measured velocity dispersions required large and
uncertain corrections for the tangential ($\sigma_\phi$) and radial
($\sigma_{\rm r}$) components of an assumed velocity dispersion
ellipsoid.

\section{The Disk Mass Project}

Measuring $\sigma_{\rm z}$ in kinematically cold stellar disks
requires spectroscopy at moderately high resolution (R$\approx$10$^4$)
of extended light at relatively low surface brightness levels
($\mu_{\rm \footnotesize B}\approx 24$ mag/arcsec$^2$).  Clearly,
measurements of $\sigma_{\rm z}$ have been severely hampered by the
limited signal-to-noise of the observations to date, as well as the
small samples of galaxies studied so far.

With the advent of Integral Field Unit (IFU) spectroscopy,
the observational situation can be dramatically improved, and
significant progress in determining the mass surface densities of
stellar disks can now be made. The main advantage of IFU spectroscopy
over traditional long-slit studies lies in the fact that IFUs are
capable to collect light from a much larger solid angle and that many
IFU spectra can be combined to increase the signal-to-noise.

Capitalizing on this aspect of IFU spectroscopy, we have initiated our
long-term Disk Mass project. The main goal is to measure $\sigma_{\rm
  z}$ as a function of radius out to 2.5 disk scale lengths in
$\sim$40 undisturbed, nearly face-on spiral galaxies with a wide range
of global properties like total luminosity, surface brightness,
colour, and morphology. To achieve this, we have constructed two
special-purpose wide-field fiber-based IFUs, and adopted a two-phased
observational strategy extending over a 3-4 year period.

\begin{figure}[t]
\begin{center}
\includegraphics[width=\textwidth]{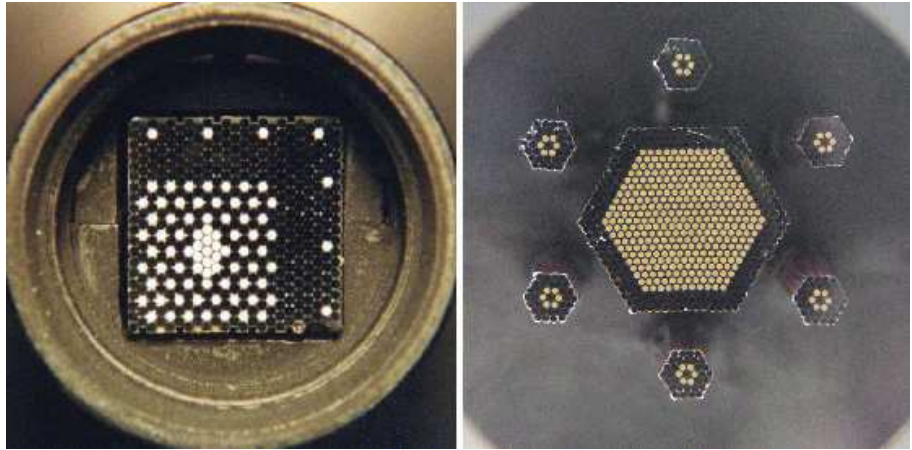}

\caption{Two custom-built fiber-based wide-field IFUs. Active fibers
  are back-illuminated. Dark fibers terminate shortly behind the focal
  plane and serve as buffer for stress relief and to edge-protect the
  active fibers while polishing the fiber head.{\bf Left:} focal plane
  layout of the fibers in the SparsePak IFU: the grid is filled with 3
  pointings. {\bf Right:} focal plane layout of the fibers in the
  P-Pak IFU. Sky fibers are located in 6 mini-IFUs surrounding the
  main fiber head.}

\end{center}
\end{figure}

Phase A aims at collecting H$\alpha$ velocity fields for a parent
sample of nearly phase-on spiral galaxies. From the UGC, we
selected disk galaxies at $|b|\ge 25^\circ$ to minimize Galactic
extinction, with diameters between 1$^\prime$ and 1.5$^\prime$ to
match them to the large field-of-view of our IFUs, and with optical
axis ratios of b/a$>$0.85 to ensure a nearly face-on orientation. This
yielded a total sample of 470 galaxies from which we removed the
strongly barred and interacting galaxies, and randomly picked galaxies
to observe.

Subsequently, the regularity of the gas kinematics is evaluated from
the H$\alpha$ velocity fields. The main purpose is to identify
kinematically disturbed disks which are likely to violate the
assumption of local isothermal equilibrium, which is required when
relating $\sigma_{\rm z}$ and disk scale height to the mass surface
density of the disk. Furthermore, Andersen \& Bershady (2003) have
demonstrated that accurate inclinations and rotation curves can be
derived for nearly face-on disks, provided a regular and symmetric
H$\alpha$ velocity field of high signal-to-noise. The parent sample
will be expanded until 40 galaxies with regular gas kinematics have
been identified.

\begin{figure}[t]
\begin{center}
\includegraphics[width=\textwidth]{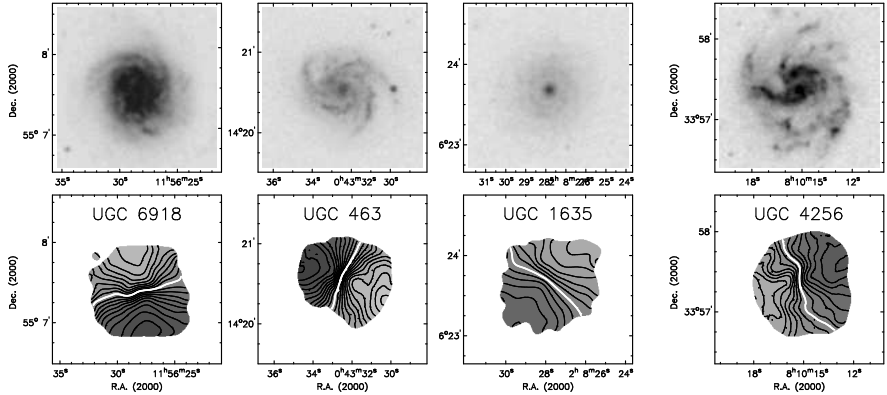}

\caption{Examples of galaxies in the parent sample. {\bf Upper row:}
optical images at the same grayscale levels. {\bf Bottom row:}
H$\alpha$ velocity fields obtained with SparsePak. The three galaxies
on the left, with different central disk surface brightness levels,
are suitable for follow-up observations of their stellar kinematics
which are presented in Figure~3. UGC 4256 is an example of a galaxy which
is too irregular, as is the case for most galaxies in our parent sample.}

\end{center}
\end{figure}

Phase B of our project aims at measuring $\sigma_{\rm z}$ in the
stellar disks of the subsample of 40 galaxies with regular gas
kinematics. Velocity dispersions are determined from the broadening of
the absorption lines in the blue part of the spectrum around 515 nm,
containing absorption lines of the MgIb triplet and many Fe lines. For
a selected number of galaxies, $\sigma_{\rm z}$ is also measured from
the broadened CaII triplet absorption lines around 860 nm.

Apart from these spectroscopic observations with our IFUs, all
galaxies in the parent sample will be imaged in the U,B,V,R,I and
J,H,K passbands. All galaxies in the subsample will be imaged in
neutral hydrogen to determine the contribution from the cold gas to
the total surface mass density of the disks.


\section{Two custom-built Integral Field Units}

As mentioned above, measuring $\sigma_{\rm z}$ requires spectroscopy
at moderately high spectral resolution of diffuse low surface
brightness light. To achieve this, we have built two special-purpose
wide-field IFUs consisting of large aperture fibers which carry light from
a focal plane to the pseudo-slit of a pre-existing spectrograph. The
diameter of the fibers is maximized while providing a spectral
resolution of no less than R$\approx$8000. The maximum number of
fibers is then determined by the length of the spectrograph slit, while the
layout in the focal plane is designed to span more than an arcminute
on the sky. Obviously, the penalty paid for large aperture fibers is a
limited angular resolution, but this is of secondary importance for
our Disk Mass project.

SparsePak, built at the University of Wisconsin in Madison, contains
75 science and 7 sky fibers, each 4.7 arcsec in diameter (Fig.~1). The
25m long fibers pipe light from a
71$^{\prime\prime}\times$72$^{\prime\prime}$ field-of-view (fov) at the F/6
imaging port of the 3.6m WIYN telescope at Kitt Peak to its Bench
Spectrograph. SparsePak is described in detail by Bershady et al
(2004, 2005).

P-Pak, built at the AIP in Potsdam, contains 331 science and 36 sky
fibers, each 2.7 arcsec in diameter (Fig.~1). The 3m long fibers carry
light from a 64$^{\prime\prime}\times$74$^{\prime\prime}$ hexagonal
fov at the focal plane behind a F/3.5 focal reducer lens on the 3.5m
Calar Alto telescope, to the collimator lens of the Cassegrain mounted
PMAS spectrograph (Verheijen et al 2004, Kelz et al 2005).  Fifteen
additional fibers allow for an accurate simultaneous wavelength
calibration.

\begin{figure}
\begin{center}
\includegraphics[width=\textwidth]{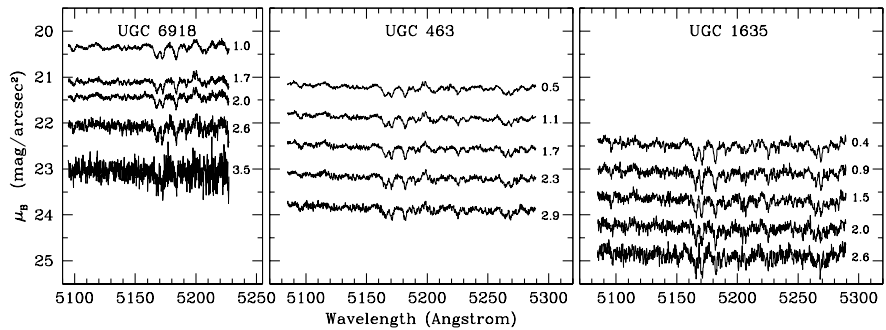}

\caption{Azimuthally averaged absorption line spectra at 5 radial bins
for 3 galaxies with different central surface brightness. Each
spectrum is plotted at the corresponding surface brightness level. At
the right of each spectrum, the number of disk scale lengths for the
radial bin is indicated. U6918 was observed for 3$\times$45$^{\rm
min}$ with SparsePak at R=11,750. U463 and U1635 were observed for
5$\times$60$^{\rm min}$ with P-Pak at R=7,800. The number of averaged
spectra at each radius is 6, 8, 12, 18, and 18 for SparsePak, and 18,
42, 66, 90, and 114 for P-Pak.}

\end{center}
\end{figure}

\section{Current status}

Phase A is effectively complete. From analyzing the gas kinematics
(Fig.~2), it became clear that only about 1 in 3 galaxies have
sufficiently regular gas kinematics to succesfully fit a tilted-ring
model to the H$\alpha$ velocity field, allowing us to measure the
shape and amplitude of the inner rotation curves from which to total
mass (dark plus luminous) of the inner galaxy follows. Hence, a parent
sample of 130 galaxies has been constructed from which a subsample of
40 regular galaxies can be selected.

We have started Phase B in the fall of 2004 and observed the stellar
kinematics in the MgIb region with P-Pak for 11 galaxies. With
SparsePak we have observed the MgIb kinematics for 8 and CaII
kinematics of 7 galaxies. With both IFUs we have built up an
extensive library of spectroscopic template stars covering a range of
spectral types, metallicities and log(g).

\begin{figure}
\begin{center}
\includegraphics[width=\textwidth]{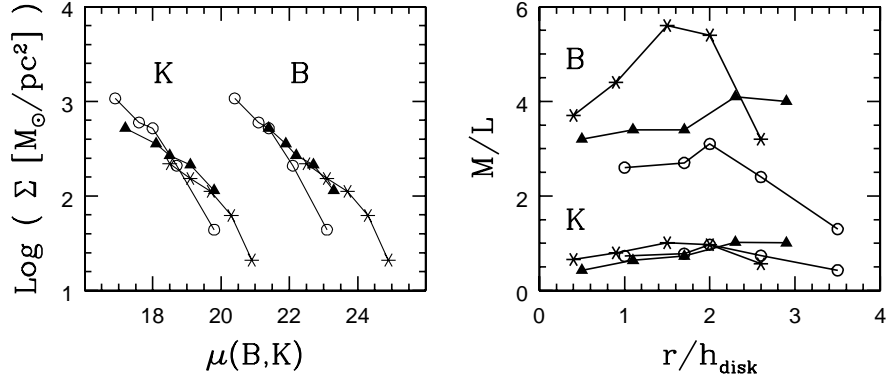}

\caption{{\bf Left:} Mass surface density as function of local B- and
K-band surface brightness. {\bf Right:} Mass-to-light ratios in the B-
and K-band as a function of radius. Circles: U6918, triangles: U463,
asteriks: U1635. Lacking K-band photometry for U1635
forced us to adopt a B$-$K colour constant with radius.}

\end{center}
\end{figure}

\section{First results}

Figure 3 shows, for 3 galaxies with different disk central surface
brightness levels, the azimuthally averaged stellar absorption line spectra
in the MgIb region of the spectra for five radial bins. Using the
stellar template spectra, we have measured $\sigma_{\rm z}$ as a
function of radius, assuming a Gaussian broadening function. Figure 4
shows the derived mass surface densities and mass-to-light
ratios. Although we are dealing with three entirely different
galaxies, the K-band M/L is quite similar for these galaxies while the
B-band M/L seems to increase systematically with lower central disk
surface brightness.

\begin{acknowledgments}

This work benefits from NSF grant AST--0307417. P-Pak was developed
within the ULTROS project, which is funded by the German ministry of
education \& research (BMBF) through Verbundforschungs grant
05--AE2BAA/4.

\end{acknowledgments}


\begin{chapthebibliography}{1}

\bibitem[Andersen \& Bershady (2003)]{and03}
Andersen, D.R. and Bershady, M.A. 2003, ApJ, 599, 79

\bibitem[Bell & de Jong (2001)]{bel01}
Bell, E.F. and de Jong, R.S. 2001, ApJ, 550, 212

\bibitem[Bershady et al (2004)]{ber04}
Bershady, M.A., Andersen, D.R., Harker, J., Ramsey, L.W. and
Verheijen, M.A.W. 2004, PASP, 116, 565

\bibitem[Bershady et al (2005)]{ber05}
Bershady, M.A., Andersen, D.R., Verheijen, M.A.W., Westfall, K.B.,
Crawford, S.M. and Swaters, R.A. 2005, ApJS, 156, 311

\bibitem[Bottema (1997)]{bot97}
Bottema, R. 1997, A\&A, 328, 517

\bibitem[de Grijs & van der Kruit (1996)]{gri96}
de Grijs, R. and van der Kruit, P.C. 1996, A\&AS, 117, 19

\bibitem[Kelz et al (2005)]{kel05}
Kelz, A., Verheijen, M.A.W., Roth, M.M., Bauer, S.M., Becker, T.,
Paschke, J., Popow, E., Sanchez, S.F. and Laux, U. 2005, PASP, in press

\bibitem[Kregel et al (2002)]{kre02}
Kregel, M., van der Kruit, P.C.and de Grijs, R. 2002, MNRAS, 334, 646

\bibitem[van Albada & Sancisi (1986)]{alb86}
van Albada, T.S. and Sancisi, R. 1886, Phil. Trans. Royal Society of
London, 320, 446

\bibitem[Verheijen et al 2004]{ver04}
Verheijen, M.A.W., Bershady, M.A., Andersen, D.R., Swaters, R.A.,
Westfall, K.B., Kelz, A. and Roth, M.M. 2004, AN, 325, 151

\end{chapthebibliography}
\end{document}